\documentclass[twocolumn,showpacs,preprintnumbers,floatfix,nofootinbib]{revtex4}
\usepackage{amssymb}
\usepackage{amsmath}
\usepackage{dcolumn}
\usepackage{bm}
\usepackage{graphicx,epsfig}
\usepackage[english]{babel}

\begin{document}

\title{What is the Criterion for a Strong First Order Electroweak Phase
Transition in Singlet Models?}
\author{Amine Ahriche}
\email{amin@physik.uni-bielefeld.de}
\affiliation{Faculty of Physics, University of Bielefeld, Postfach 100131, D-33501
Bielefeld, Germany.\\
Department of Physics, University of Jijel, BP 98, Ouled Aissa, DZ-18000
Jijel, Algeria.}

\begin{abstract}
It is widely believed that the existence of singlet scalars in some Standard
Model extensions can easily make the electroweak phase transition strongly
first order, which is needed for the electroweak baryogenesis scenario. In
this paper, we will examine the strength of the electroweak phase transition
in the simplest extension of the Standard Model with a real singlet using
the sphaleron energy at the critical temperature. We find that the phase
transition is stronger by adding a singlet; and also that the criterion for
a strong phase transition $\Omega (T_{c})/T_{c}\gtrsim 1$, where $\Omega
=(\upsilon ^{2}+(x-x_{0})^{2})^{\frac{1}{2}}$ and $x$ ($x_{0}$) is the
singlet vev in the broken (symmetric) phase, is not valid for models
containing singlets, even though often used in the literature. The usual
condition $\upsilon _{c}/T_{c}\gtrsim 1$ is more meaningful, and it is
satisfied for a large part of the parameter space for physically allowed
Higgs masses.
\end{abstract}

\pacs{98.80.Cq  11.10.Wx  11.15.Ha}
\maketitle

\section{Introduction}

The Standard Cosmological Model has been successful in describing the early
universe, it is supported by a number of important observations: the
expansion of the universe, the abundance of the light elements and the
cosmic microwave background (CMB) radiation. These three measurable
signatures strongly support the notion that our universe evolved from a
dense, nearly featureless hot gas, just as the Big Bang model predicts.

However it fails to explain some serious problems like the nature of dark
matter and dark energy; and the dominance of matter over antimatter.

From a theoretical point of view, there is no justification to assume that
the universe started its evolution with a defined baryon asymmetry; $%
n_{b}\left( t=0\right) >n_{\bar{b}}\left( t=0\right) $. The natural
assumption is that the universe was initially neutral. Direct observations
show that the universe contains no appreciable primordial antimatter. In
addition, the success of big bang nucleosynthesis requires that the ratio of
the effective baryon number ($n_{b}-n_{\bar{b}}$) to the entropy density
should be between \cite{BBN}
\begin{equation}
2.6\times 10^{-10}<\eta \equiv \frac{n_{b}-n_{\bar{b}}}{s}<6.2\times
10^{-10}.  \label{eta}
\end{equation}
This number has been independently determined to be $\eta =(8.7\pm
0.3)\times 10^{-11}$ from precise measurements of the relative heights of
the first two microwave background ($CMB$) acoustic peaks by the WMAP
satellite \cite{wmap}. Thus how can one understand the origin of this
asymmetry? This is what is called the problem of baryogenesis (for a review
see \cite{rev}). In 1967, Sakharov forwarded his three conditions for
baryogenesis \cite{sak}, which are summarized in the existence of processes
which: $(1)$ violate $B$ number, $(2)$ violate $C$ and $CP$ symmetries; and $%
(3)$ take place out of equilibrium.

One of the most interesting scenarios for baryogenesis is the electroweak
baryogenesis (For a review, see \cite{ewbar}), where the third Sakharov
condition is realized via a strong first order phase transition at the
electroweak scale.

In gauge theories, a first order phase transition takes place if the vacuum
of the theory does not correspond to the global minimum of the potential.
Since it is energetically unfavored, the field changes its value to the true
vacuum (i.e. the absolute minimum of the potential). Because of the
existence of a barrier between the two minima, this mechanism can happen by
tunneling or thermal fluctuations via bubble nucleation. The electroweak
baryogenesis scenario is realized when the $B$ and $CP$ violating
interactions pass through the bubble wall. These interactions are very fast
outside the bubbles but suppressed inside. Then a net baryon asymmetry
results inside the bubbles which are expanding and filling the universe at
the end.

In order to compute the net baryon number, the rate of $B$ violating
processes in the broken phase is needed. In the Standard Model, $B$ number
is violated at the quantum level \cite{thooft}, where the transition between
two topologically distinct $SU(2)_{L}$ ground states is possible.

The transition between two neighboring ground states breaks both lepton and
baryon numbers by $\Delta L=\Delta B=3$. To find the rate of $B$ violating
processes, one needs to know the sphaleron solution, which is a static field
configuration that interpolates between the two distinct ground states. The
sphaleron configuration was found in \cite{sphaleron} for the $SU(2)_{L}$
model.

A model-independent condition in order that the phase transition be strong
enough was derived in \cite{cond}:
\begin{equation}
E_{sp}\left( T_{c}\right) /T_{c}>45,  \label{con1}
\end{equation}
where $E_{sp}$ and $T_{c}$ are the sphaleron energy and the critical
temperature, respectively. Since it was shown in~\cite{SMcheck} that $%
E_{Sp}\left( T\right) \varpropto \upsilon \left( T\right)$ \footnote{%
This was also checked for the Minimal Supersymmetric Standard Model (MSSM)
in \cite{MSSMcheck}, then (\ref{con}) is valid also for the MSSM.}, the
condition (\ref{con1}) can be translated for the case of Standard Model to
\cite{phicond}%
\begin{equation}
\upsilon _{c}/T_{c}>1,  \label{con}
\end{equation}%
where $\upsilon _{c}$\ is the field value at the critical temperature.
However this condition (\ref{con}) is not fulfilled in the case of Standard
Model, because the thermal induced cubic term\footnote{%
This term is forbidden by symmetry at tree level, however it appears as $%
T(m^{2})^{\frac{3}{2}}$ from the thermal bosonic contributions to the
effective potential at one-loop.} is not large enough; also this leads to an
unacceptable upper bound on the Higgs mass \cite{mhbound}
\begin{equation}
m_{h}\leq 42~GeV.  \label{mhbound}
\end{equation}

The constraint gets even stronger when the two-loops effect and a proper
treatment of the top quark are included \cite{Heb}. It is clear that this
bound is in contradiction with the lower bound coming from LEP $m_{h}>114$ $%
GeV$ \cite{LEP}.

But this severe bound can be avoided by adding a complex scalar singlet that
couples only to the Higgs doublet with an appropriate choice of the theory
parameters in a way that the singlet does not develop a vev \cite{zvs}.

If a new scalar (or many scalars which can be singlets or in doublet w.r.t $%
SU(2)_{L}$) acquiring a vacuum expectation value $x$\ are added to the
Standard Model, the term $\upsilon _{c}$\ in (\ref{con}) should perhaps be
replaced by $\Omega _{c}$ which equals $\{\upsilon _{c}^{2}+x_{c}^{2}\}^{%
\frac{1}{2}}$;\ or $\{\upsilon _{c}^{2}+\left( x-x_{0}\right) _{c}^{2}\}^{%
\frac{1}{2}}$\ when the false vacuum is $(0,x_{0})$ instead of $(0,0)$ \cite%
{SmS}. Then (\ref{con}) becomes%
\begin{equation}
\Omega _{c}/T_{c}>1.  \label{con2}
\end{equation}

If the new particle(s) is a singlet(s), cubic terms can exist in the
potential at tree-level, and therefore the phase transition gets stronger
without the need of the thermally induced one \cite{SmS,SmS0}.

In this work, we want to check for a model with a singlet whether the
passage from (\ref{con1}) to (\ref{con2}) is true or not? We will consider
the simplest extension of the Standard Model with a real singlet. This paper
is organized as follows: In the second section, we introduce briefly this
model, and find the sphaleron solution in the third section. In the fourth
section, we discuss the strength of the first order electroweak phase
transition (EWPT). And finally we give our conclusion.

\section{The Standard Model with a Singlet 'SM+S'}

Let us consider an extension of the Standard Model by a singlet real scalar $%
S$ coupled only to the standard Higgs. We concentrate here on the scalar
sector (SM Higgs and the added singlet) and the $SU(2)_{L}$ gauge sector.%
\footnote{%
Since we are interested here in the Sphaleron solution, we assume that the $%
U_{Y}(1)$ contribution to the sphaleron energy to be negligible as in the
case of Standard Model \cite{fmixing}.}

\begin{center}
\textbf{The effective Lagrangian}
\end{center}

The Lagrangian is given by%
\begin{eqnarray}
\mathcal{L} &=&-\tfrac{1}{4}F_{\mu \nu }^{a}F^{a\mu \nu }+\left( D_{\mu
}\phi \right) ^{\dagger }\left( D^{\mu }\phi \right) +\tfrac{1}{2}\left(
\partial _{\mu }S\right) \left( \partial ^{\mu }S\right)  \notag \\
&&-V_{eff}\left( \phi ,S\right) ,  \label{Lagr}
\end{eqnarray}%
where $\phi $ is the Higgs doublet%
\begin{equation}
\phi ^{T}=1/\sqrt{2}\left(
\begin{array}{cc}
\chi _{1}+i\chi _{2} & h+i\chi _{3}%
\end{array}%
\right)
\end{equation}%
where $h$ is the scalar standard Higgs, $\chi $'s\ are the three Goldstone
bosons, and $F_{\mu \nu }^{a}$\ is the $SU(2)_{L}$ field strength
\begin{equation}
F_{\mu \nu }^{a}=\partial _{\mu }A_{\nu }^{a}-\partial _{\nu }A_{\mu
}^{a}+g\epsilon ^{abc}A_{\mu }^{b}A_{\nu }^{c}.
\end{equation}%
$D_{\mu }$\ is the covariant derivative; when neglecting the $U_{Y}(1)$
gauge group, it is given by
\begin{equation}
D_{\mu }=\partial _{\mu }-\tfrac{i}{2}g\sigma ^{a}A_{\mu }^{a}.
\end{equation}%
Finally, $V_{eff}\left( \phi ,S\right) $\ is the effective potential, which
is at tree-level given by%
\begin{eqnarray}
V_{0}\left( \phi ,S\right) &=&\lambda \left\vert \phi \right\vert ^{4}-\mu
_{h}^{2}\left\vert \phi \right\vert ^{2}+\omega \left\vert \phi \right\vert
^{2}S^{2}+\rho \left\vert \phi \right\vert ^{2}S  \notag \\
&&+\tfrac{\lambda _{S}}{4}S^{4}-\tfrac{\alpha }{3}S^{3}-\tfrac{\mu _{S}^{2}}{%
2}S^{2}.  \label{tree}
\end{eqnarray}

We can eliminate $\mu _{h}^{2}$ and $\mu _{S}^{2}$\ by making $\left(
\upsilon ,x\right) $ as the absolute minimum of\ the one-loop effective
potential at zero temperature, where $\upsilon =246.22$ GeV is the standard
Higgs vev.

Now, we write the explicit formula of the one-loop effective potential. We
will consider the contributions of the gauge bosons, the standard Higgs $h$,
the singlet $S$, the Goldstone bosons $\chi _{1,2,3}$ and the top quark. The
field-dependent masses at zero temperature are given by%
\begin{gather}
m_{t}^{2}=\tfrac{1}{2}y_{t}^{2}h^{2},\text{ }m_{Z}^{2}=\tfrac{\bar{g}%
^{2}+g^{^{\prime }2}}{4}h^{2},~m_{W}^{2}=\tfrac{g^{2}}{4}h^{2},  \notag \\
m_{\chi }^{2}=\lambda h^{2}-\mu _{h}^{2}+\omega S^{2}+\rho S  \notag \\
m_{h,S}^{2}\rightarrow m_{1,2}^{2}=\tfrac{1}{2}[\left( 3\lambda +\omega
\right) h^{2}+\left( 3\lambda _{S}+\omega \right) S^{2}  \notag \\
+\left( \rho -2\alpha \right) S-\mu _{h}^{2}-\mu _{S}^{2}  \notag \\
\mp \left\{ \left( \left( 3\lambda -\omega \right) h^{2}-\left( 3\lambda
_{S}-\omega \right) S^{2}+\left( \rho +2\alpha \right) S-\mu _{h}^{2}+\mu
_{S}^{2}\right) ^{2}\right.  \notag \\
\left. +4\left( 2\omega S+\rho\right) ^{2}h^{2}\right\} ^{\frac{1}{2}}]
\label{allmasses}
\end{gather}%
where $y_{t}$ is the Yukawa coupling for the top quark, and $\bar{g}%
^{2}=g^{2}+g^{^{\prime }2}$, however we neglected the $U_{Y}(1)$ gauge group
and therefore $g^{^{\prime }}=0$ and $m_{Z}=m_{W}$; and $m_{1,2}^{2}$\ are
the Higgs-singlet eigenmasses. Then the one-loop correction to the effective
potential at zero temperature is given by%
\begin{eqnarray}
V_{1}^{T=0}\left( h,S\right) &=&\sum_{i=W,Z,t,h,S,\chi }n_{i}G\left(
m_{i}^{2}\left( h,S\right) \right)  \label{1lz} \\
\text{and }G\left( x\right) &=&\frac{x^{2}}{64\pi ^{2}}\left\{ \log \left(
\frac{x}{Q^{2}}\right) -\frac{3}{2}\right\} .  \notag
\end{eqnarray}%
Here $Q$ is the renormalization scale, which we take to be the standard
Higgs vev $Q=\upsilon $; and $n_{i}$\ are the particle degree of freedom;
which are%
\begin{equation}
n_{W}=6,n_{Z}=3,n_{h}=1,n_{\chi }=3,n_{S}=1,n_{t}=-12.
\end{equation}

The temperature-dependent part at one loop is given by \cite{thermal}%
\begin{equation}
V_{1}^{T\neq 0}\left( h,S,T\right) =\frac{T^{4}}{2\pi ^{2}}%
\sum_{i=W,Z,t,h,S,\chi }n_{i}J_{B,F}\left( m_{i}^{2}(h,S)/T^{2}\right)
\label{1lf}
\end{equation}%
and%
\begin{eqnarray}
J_{B}\left( \theta \right) &=&\int_{0}^{\infty }dx~x^{2}\log \left\{ 1-\exp
\left[ -\sqrt{x^{2}+\theta }\right] \right\}  \notag \\
&&\overset{\theta \ll 1}{\simeq }-\tfrac{\pi ^{4}}{45}+\tfrac{\pi ^{2}}{12}%
\theta -\tfrac{\pi }{6}\theta ^{3/2}-\tfrac{\theta ^{2}}{32}\log \tfrac{%
\theta }{a_{b}},  \label{bos} \\
J_{F}\left( \theta \right) &=&\int_{0}^{\infty }dx~x^{2}\log \left\{ 1+\exp
\left[ -\sqrt{x^{2}+\theta }\right] \right\}  \notag \\
&&\overset{\theta \ll 1}{\simeq }\tfrac{7\pi ^{4}}{360}-\tfrac{\pi ^{2}}{24}%
\theta -\tfrac{\theta ^{2}}{32}\log \tfrac{\theta }{a_{f}},  \label{fer}
\end{eqnarray}%
where $a_{b}=16\pi ^{2}\exp (3/2-2\gamma _{E})$, $a_{f}=\pi ^{2}\exp
(3/2-2\gamma _{E})$ and $\gamma _{E}=0.5772156649$ is the Euler constant.
There is also another part of the effective potential which is the \textit{%
ring }(or\textit{\ daisy}) contribution \cite{ring}%
\begin{eqnarray}
V_{ring}\left( h,S,T\right) &=&-\tfrac{T}{12\pi }\sum_{i=W,Z,h,S,\chi
}n_{i}\left\{ \left( M_{i}^{2}\left( h,S,T\right) \right) ^{\frac{3}{2}%
}\right.  \notag \\
&&\left. -\left( m_{i}^{2}\left( h,S\right) \right) ^{\frac{3}{2}}\right\} ,
\label{daisy}
\end{eqnarray}%
where $M_{i}^{2}\left( h,S,T\right) $'s\ are the thermal masses of the
bosons, which are given by%
\begin{equation}
M_{i}^{2}\left( h,S,T\right) =m_{i}^{2}\left( h,S\right) +\Pi _{i}
\end{equation}%
and $\Pi _{i}$\ is the thermal correction to the mass, its values for the
bosons in our model are:
\begin{eqnarray}
\Pi _{W}^{L} &=&\frac{11}{6}g^{2}T^{2},~\Pi _{W}^{T}=0  \notag \\
\Pi _{Z}^{L} &=&\frac{11}{6}g^{2}T^{2},~\Pi _{Z}^{T}=0  \notag \\
\Pi _{\chi } &=&\left( g^{2}/4+\lambda /2+y_{t}^{2}/4+\omega /6\right) T^{2}
\notag \\
\Pi _{hh} &=&\left( g^{2}/4+3\lambda /2+y_{t}^{2}/4+\omega /6\right) T^{2}
\notag \\
\Pi _{SS} &=&\left( \lambda _{S}/4+2\omega /3\right) T^{2},~\Pi _{hS}\simeq 0
\end{eqnarray}%
where the script $L$ ($T$) denotes the longitudinal (transversal) mode for $%
W $ and $Z$. Then the full one-loop effective potential at finite
temperature is the sum of (\ref{tree}), (\ref{1lz}), (\ref{1lf}) and (\ref%
{daisy}):%
\begin{gather}
V_{eff}^{1-loop}\left( h,S,T\right) =V_{0}\left( h,S\right)
+\sum_{i=W,Z,t,h,S,\chi }n_{i}G\left( m_{i}^{2}\left( h,S\right) \right)
\notag \\
+\frac{T^{4}}{2\pi ^{2}}\sum_{i=W,Z,t,h,S,\chi }n_{i}J_{B,F}\left(
m_{i}^{2}(h,S)/T^{2}\right)  \notag \\
-\tfrac{T}{12\pi }\sum_{i=W,Z,h,S,\chi }n_{i}\left\{ \left( M_{i}^{2}\left(
h,S,T\right) \right) ^{\frac{3}{2}}-\left( m_{i}^{2}\left( h,S\right)
\right) ^{\frac{3}{2}}\right\} .  \label{1ltf}
\end{gather}

The mass-squared values of the Goldstone bosons or the Higgs-singlet
eigenstates can be negative. In the case where a mass value (or more) is
negative, the cubic term in (\ref{bos}) becomes non analytic, that's not a
problem since it is already replaced by the thermal mass in (\ref{daisy}),
where it will be compensated by the thermal correction. If the thermal
correction is not enough to compensate the negative value, this term should
be omitted since it is imaginary and does not belong to the effective
potential.

\begin{center}
\textbf{The space of parameters}
\end{center}

In our theory, we have quite a few parameters,
\begin{equation*}
\lambda ,~\lambda _{S},~\omega ,~\rho ,~\alpha ,~\mu _{h}^{2},~\mu _{S}^{2},
\end{equation*}
in addition to the singlet vev $x$. As mentioned above, $\mu _{h}^{2}$\ and $%
\mu _{S}^{2}$\ can be eliminated as
\begin{eqnarray}
\mu _{h}^{2} &=&\lambda \upsilon ^{2}+\omega x^{2}+\rho x+\tfrac{1}{\upsilon
}\left. \tfrac{\partial V_{1}^{T=0}\left( h,S\right) }{\partial h}%
\right\vert _{\substack{ h=\upsilon  \\ S=x}} \\
\mu _{S}^{2} &=&\omega \upsilon ^{2}+\tfrac{\rho \upsilon ^{2}}{2x}+\lambda
_{S}x^{2}-\alpha x+\tfrac{1}{x}\left. \tfrac{\partial V_{1}^{T=0}\left(
h,S\right) }{\partial S}\right\vert _{\substack{ h=\upsilon  \\ S=x}}
\end{eqnarray}
after which our free parameters are $\lambda ,~\lambda _{S},~\omega ,~\rho
,~\alpha ~$and $x$. Since the theory is invariant under the discrete
symmetry ($x$,$\rho $,$\alpha $)$\rightarrow $(-$x$,-$\rho $,-$\alpha $), we
will assume only positive values for the singlet vev $x$. We want also to
keep the perturbativity of theory by imposing $\lambda ,\lambda
_{S},\left\vert \omega \right\vert \ll 1$. We choose the parameters, $%
\lambda ,~\lambda _{S},~\omega ,~\rho ,~\alpha ~$and $x$, lying in the
ranges:%
\begin{eqnarray}
0.001 &\leq &\lambda ,\lambda _{S}\leq 0.6  \notag \\
-0.6 &\leq &\omega \leq 0.6  \notag \\
100 &\leq &x/GeV\leq 350  \notag \\
-350 &\leq &\alpha /GeV\leq 350  \notag \\
-350 &\leq &\rho /GeV\leq 350  \label{pararang}
\end{eqnarray}

The stability of the theory implies that the potential goes to infinity when
the field goes to the infinity in any direction, which implies $\omega
^{2}<\lambda \times \lambda _{S}$. Moreover, we need that any minimum or
extremum of the potential should be in the range of the electroweak theory;
let us say that all the minima and extrema must be inside the circle $%
h^{2}+S^{2}=\{600$ $GeV\}^{2}$ in the $h-S$ plane; and therefore the
potential is monotonically increasing outside this circle in any direction.

In the Standard Model, the Higgs mass lower bound is given by $%
m_{h}^{SM}>114 $ $GeV$ \cite{LEP}. The mixing between the standard Higgs and
the singlet changes the couplings of the standard Higgs to all the SM sector
(gauge bosons and leptons), and therefore this bound is not viable. In our
work, we will not derive the new lower bound for the Higgs mass, but we will
restrict ourselves only with masses $m_{1,2}$ in the range $65$ $GeV$ to $%
450 $ $GeV$.

\section{Sphaleron in the 'SM+S'}

In order to find the sphaleron solution for this model, we follow the same
steps as in the $SU(2)_{L}$ model. Applying Euler-Lagrange conditions on the
effective Lagrangian, (\ref{1lz}) or (\ref{1ltf}), we find the field
equations%
\begin{gather}
\partial _{\gamma }F^{q\gamma \tau }-g\epsilon ^{qab}A_{\alpha
}^{b}F^{a\alpha \tau }+\tfrac{1}{4}g^{2}h^{2}A^{q\tau }=0  \notag \\
\partial ^{2}h-\tfrac{1}{4}g^{2}hA_{\mu }^{a}A^{a\mu }+\tfrac{\partial }{%
\partial h}V_{eff}\left( \phi ,S,T\right) =0  \notag \\
\partial ^{2}S+\tfrac{\partial }{\partial S}V_{eff}\left( \phi ,S,T\right)
=0.  \label{eqsms}
\end{gather}

We will work in the orthogonal gauge where%
\begin{equation}
A_{0}=0,~~x_{i} A_{i}=0.
\end{equation}

We will not use the spherically symmetric ansatz for \{$\phi ,A_{i}$\} in
\cite{sphaleron}, but another equivalent one \cite{spmod},%
\begin{eqnarray}
A_{i}^{a}\left( x\right) &=&2\left( 1-f\left( r\right) \right) \dfrac{%
\epsilon _{aij}x_{j}}{gr^{2}}  \notag \\
H\left( x\right) &=&\tfrac{h}{\sqrt{2}}\left(
\begin{array}{c}
0 \\
1%
\end{array}%
\right) ,~~h=\upsilon L\left( r\right)  \notag \\
S\left( x\right) &=&xR\left( r\right) .
\end{eqnarray}%
Here $\upsilon $\ and $x$\ are the Higgs and singlet vevs in the general
case (zero or nonzero temperature). Then one can rewrite the field equations
(\ref{eqsms}) as\footnote{%
There is a similar work done in \cite{choi}, however there is a difference
in the definition\ in the theory parameters, and also there is an error on
the r.h.s of the first equation in (19) in this paper, where the term $%
u^{2}/\upsilon ^{2}$ should be corrected as $u^{2}/V^{2}$ according to his
notation. In our notation it is the term $\upsilon ^{2}/\Omega ^{2}$ in the
first equation in (\ref{DifEq}).}%
\begin{eqnarray}
\zeta ^{2}\frac{\partial ^{2}}{\partial \zeta ^{2}}f &=&2f\left( 1-f\right)
\left( 1-2f\right) -\tfrac{1}{4}\frac{\upsilon ^{2}}{\Omega ^{2}}\zeta
^{2}L^{2}\left( 1-f\right)  \notag \\
\frac{\partial }{\partial \zeta }\zeta ^{2}\frac{\partial }{\partial \zeta }%
L &=&2L\left( 1-f\right) ^{2}+\frac{\zeta ^{2}}{g^{2}\upsilon \Omega ^{2}}%
\left. \tfrac{\partial V_{eff}\left( h,S,T\right) }{\partial h}\right\vert
_{h=\upsilon L,S=xR}  \notag \\
\frac{\partial }{\partial \zeta }\zeta ^{2}\frac{\partial }{\partial \zeta }%
R &=&\frac{\zeta ^{2}}{g^{2}x\Omega ^{2}}\left. \tfrac{\partial
V_{eff}\left( h,S,T\right) }{\partial S}\right\vert _{h=\upsilon L,S=xR}
\label{DifEq}
\end{eqnarray}%
where $\zeta =g\Omega r$; the parameter $\Omega $\ can take any
non-vanishing value of mass dimension one (for example $\upsilon $, $x$ or $%
\sqrt{\upsilon ^{2}+x^{2}}$); and the energy functional is given by%
\begin{eqnarray}
E_{Sp}\left( T\right) &=&\frac{4\pi \Omega }{g}\int_{0}^{+\infty }d\zeta
\left\{ 4\left( \frac{\partial }{\partial \zeta }f\right) ^{2}+\frac{8}{%
\zeta ^{2}}f^{2}\left( 1-f\right) ^{2}\right.  \notag \\
&&+\tfrac{1}{2}\frac{\upsilon ^{2}}{\Omega ^{2}}\zeta ^{2}\left( \frac{%
\partial }{\partial \zeta }L\right) ^{2}+\frac{\upsilon ^{2}}{\Omega ^{2}}%
L^{2}\left( 1-f\right) ^{2}  \notag \\
&&+\tfrac{1}{2}\frac{x^{2}}{\Omega ^{2}}\zeta ^{2}\left( \frac{\partial }{%
\partial \zeta }R\right) ^{2}+\frac{\zeta ^{2}}{g^{2}\Omega ^{4}}\times
\notag \\
&&\left. \left\{ V_{eff}\left( \upsilon L,xR,T\right) -V_{eff}\left(
\upsilon ,x,T\right) \right\} \right\} ,  \label{enfuncs}
\end{eqnarray}%
with the boundary conditions (See Appendix A)%
\begin{equation}
\begin{array}{cccccccc}
\text{for }\zeta \sim 0 &  & f\sim \zeta ^{2} &  &  & \text{for }\zeta
\rightarrow \infty &  & f\rightarrow 1 \\
&  & L\sim \zeta &  &  &  &  & L\rightarrow 1 \\
&  & R\sim a+b\zeta ^{2}; &  &  &  &  & R\rightarrow 1.%
\end{array}
\label{bc}
\end{equation}

Let us now compare the energy functional (\ref{enfuncs}) to that of the
minimal Standard Model (eq. (10) in \cite{sphaleron}). The difference
between these quantities is of course the contribution of the singlet, which
contains the kinetic term, the mixing with the standard Higgs; and a
contribution to the potential term. However if we compare (\ref{enfuncs})
with the same quantity in the MSSM case (Ea. (2.22) in \cite{MSSMcheck}), we
find that in the MSSM both Higgs fields, $h_{1}$ and $h_{2}$, have similar
contributions to the sphaleron energy, and its general form remains
invariant under $h_{1}\leftrightarrow h_{2}$. However this is not the case
for (\ref{enfuncs}) if $h\leftrightarrow S$, because of a missing term like $%
R^{2}\left( 1-f\right) ^{2}$. \footnote{%
To be more precise, the absence of a mixing between the singlet and the
gauge field is not the only reason to spoil this invariance, but this
invariance is absent also in the tree-level potential.}

For the MSSM sphaleron energy, its form is invariant under $%
h_{1}\leftrightarrow h_{2}$, and it scales like $\{\upsilon
_{1}^{2}+\upsilon _{2}^{2}\}^{\frac{1}{2}}$; and for our model 'SM+S', a
similar invariance is absent. Could it nevertheless be that $%
E_{Sp}\varpropto \{\upsilon ^{2}+\left( x-x_{0}\right) ^{2}\}^{\frac{1}{2}}$%
? We will check this in the next section.

But when comparing (\ref{enfuncs}) with the same quantity for the
Next-to-Supersymmetric Standard Model (NMSSM); (eq. (2.20) in \cite{SpNMSSM}%
; after eliminating explicit CP phases), we find no large difference expect
for what comes from the fact that the NMSSM contains a doublet more than the
'SM+S'; and we remark also similar equations of motion and also similar
boundary conditions.

The analytic solution of the system (\ref{DifEq}) is not possible, this
should be done numerically. To solve this system numerically, we need to
transform it into a system of 6 first-order differential equations, and
therefore we have a first order two-point boundary problem, then we use the
so-called relaxation method to solve it. This method is well explained in
section 17.3 of \cite{nr}.

As an example, we solve the system (\ref{DifEq}) for four chosen sets of
parameters (A, B, C and D); and then we can compute the sphaleron energy (%
\ref{enfuncs}) at any temperature $T\leq T_{c}$. All the results for the
sets A, B, C and D are summarized in table \ref{tab1}.

\begin{center}
\begin{table}[h]
\begin{center}
\begin{tabular}{c|c|c|c|c|}
\cline{2-5}
& A & B & C & D \\ \hline
\multicolumn{1}{|c|}{$\lambda $} & 0.4000 & 0.4000 & 0.5000 & 0.4150 \\
\hline
\multicolumn{1}{|c|}{$\lambda _{S}$} & 0.4003 & 0.4200 & 0.4100 & 0.5500 \\
\hline
\multicolumn{1}{|c|}{$\omega $} & 0.3818 & 0.2818 & 0.3818 & 0.3000 \\ \hline
\multicolumn{1}{|c|}{$x/GeV$} & 200 & 250 & 350 & 350 \\ \hline
\multicolumn{1}{|c|}{$\alpha /GeV$} & -38.89 & 38.89 & 38.89 & 194.44 \\
\hline
\multicolumn{1}{|c|}{$\rho /GeV$} & -272.22 & -194.44 & -272.22 & -300 \\
\hline
\multicolumn{1}{|c|}{$m_{1}/GeV$} & 178.00 & 204.00 & 244.74 & 203.05 \\
\hline
\multicolumn{1}{|c|}{$m_{2}/GeV$} & 311.92 & 269.80 & 333.96 & 318.80 \\
\hline
\multicolumn{1}{|c|}{$T_{c}/GeV$} & 141.55 & 241.34 & 389.94 & 270.08 \\
\hline
\multicolumn{1}{|c|}{$E_{sp}(0)/GeV$} & 9618.6 & 9721.3 & 9883.3 & 9726.6 \\
\hline
\multicolumn{1}{|c|}{$\upsilon _{c}/T_{c}$} & 1.680 & 0.838 & 0.495 & 0.386
\\ \hline
\multicolumn{1}{|c|}{$\Omega _{c}/T_{c}$} & 3.138 & 1.232 & 1.436 & 0.703 \\
\hline
\multicolumn{1}{|c|}{$E_{Sp}(T_{c})/T_{c}$} & 64.851 & 32.980 & 20.459 &
13.577 \\ \hline
\end{tabular}%
\end{center}
\caption{\textit{Representative parameter values and the corresponding
values of the scalar masses, critical temperature and different ratios
needed for the criterion of a strong first order phase transition.}}
\label{tab1}
\end{table}
\end{center}

From table \ref{tab1}, the set (A) satisfies both conditions (\ref{con2})
and (\ref{con1}), (D) does not satisfy either of them, and both (B) and (C)
satisfy (\ref{con2}) but not (\ref{con1}).

\begin{figure}[h]
\begin{center}
\includegraphics[width=9cm,height=7cm]{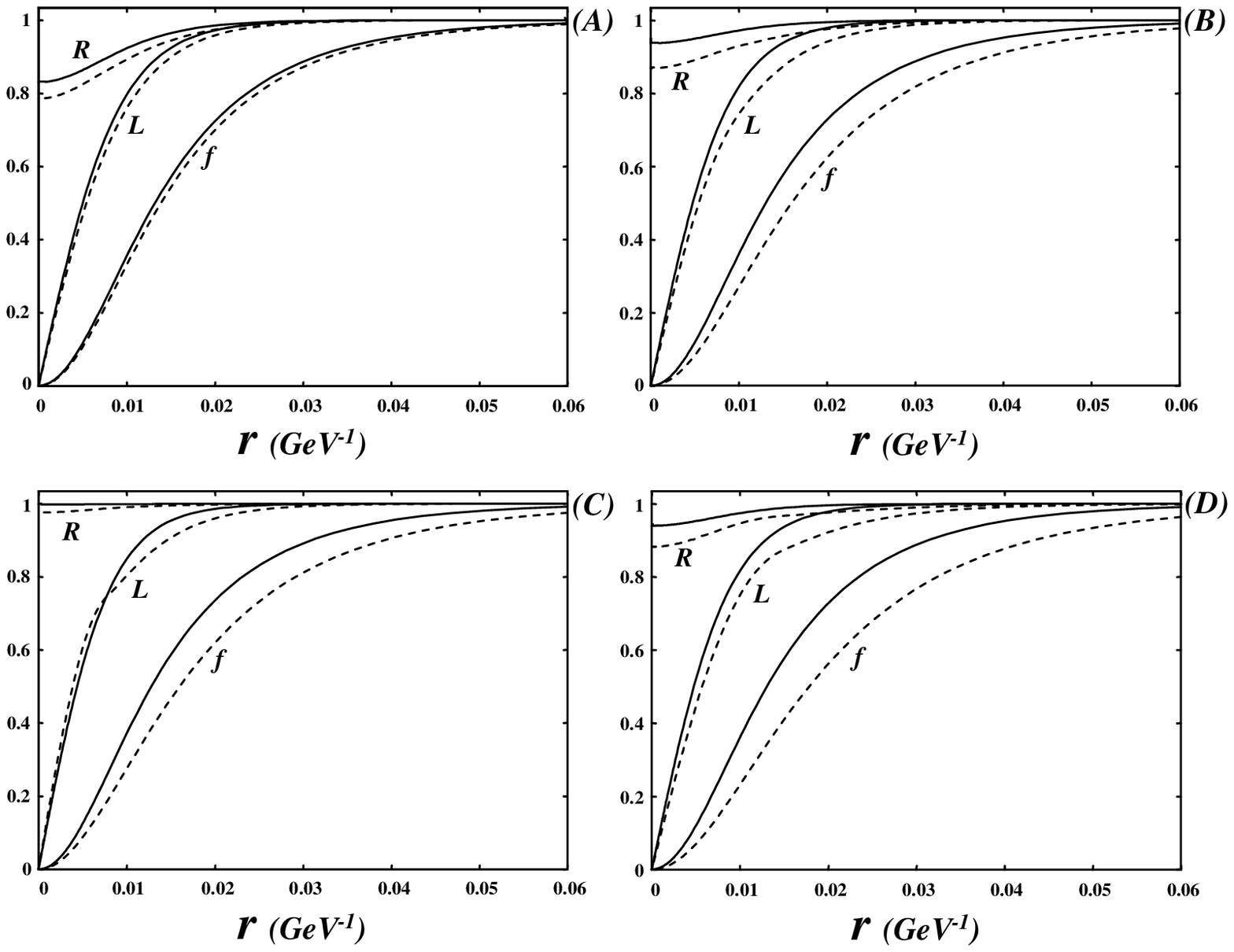}
\end{center}
\caption{\textit{A, B, C and D represent the profiles of the functions f, L
and R for the sets of parameters A, B, C and D in table \protect\ref{tab1}
respectively. The continuous lines represent the profiles at zero
temperature and the dashed ones represent the profiles of the functions at
finite temperature.}}
\label{1}
\end{figure}

The profiles of the functions $f$, $L$ and $R$ are shown in Fig. \ref{1}.

From all cases in Fig. \ref{1}, we remark that the singlet profile is not
much different than the unity, due to Neumann type boundary at $r=0$. This
comes from the fact that the singlet couples to the Higgs doublet and not
the gauge fields. Then we claim that the singlet contribution to the
sphaleron energy (\ref{enfuncs}), should be small compared to doublet and
gauge field contributions.

\section{The Phase Transition in the 'SM+S'}

In Ref \cite{SmS}, the authors have studied the EWPT strength using the same
tree-level potential as (\ref{tree}) with some differences in the parameter
definitions. They easily got a strong first order phase transition even for
Higgs masses much larger than (\ref{mhbound}). And of course they used the
criterion (\ref{con2}) instead of (\ref{con}), where the quantity $\upsilon
_{c}$\ is replaced by $\Omega _{c}=\{\upsilon _{c}^{2}+\left(
x_{c}-x_{0c}\right) ^{2}\}^{\frac{1}{2}}$.\ Since $\Omega _{c}/T_{c}\geq
\upsilon _{c}/T_{c}$\ is always fulfilled, the phase transition gets
stronger for a larger parameter space compared with the minimal Standard
Model case.

\begin{figure}[h]
\begin{center}
\includegraphics[width=7.5cm,height=11cm]{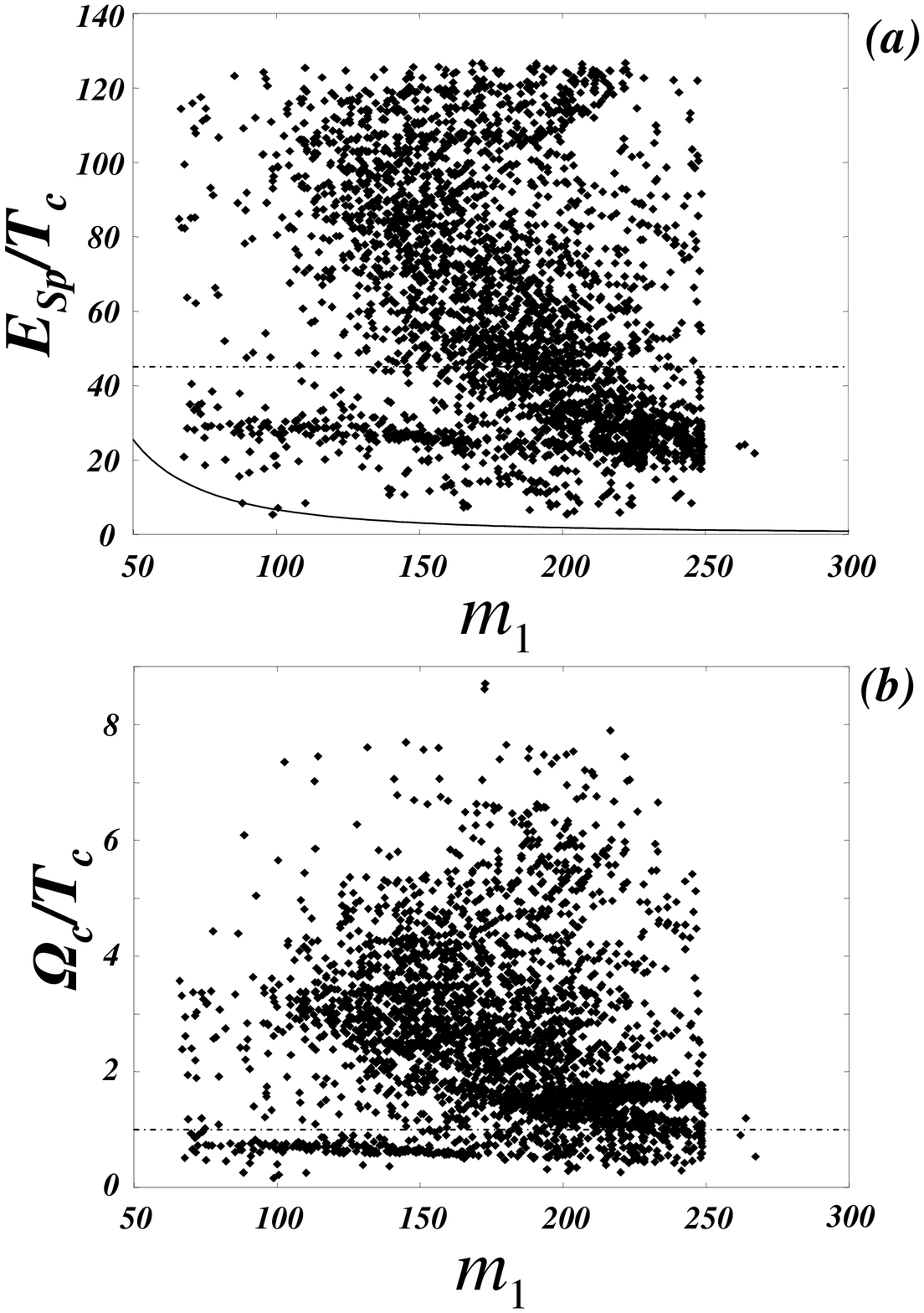}
\end{center}
\caption{\textit{For the points above the dash-dotted lines in (a) and (b),
the electroweak phase transition is strongly first order according to (
\protect\ref{con1}) and (\protect\ref{con2}), respectively. In (a), the
continuous curve represents }$\mathit{E}_{\mathit{SP}}\mathit{(T}_{\mathit{c}%
}\mathit{\ )/T}_{\mathit{c}}$\textit{\ as a function of the Higgs mass for
the case of the Standard Model.}}
\label{2}
\end{figure}

Let us take a random choice of about 3000 parameters in the ranges (\ref%
{pararang}), and make a comparison between the two different criteria of the
strong first order phase transition (\ref{con2})\ and (\ref{con1}). We show
the plots of the quantities $\Omega _{c}/T_{c}$\ and $E_{Sp}\left(
T_{c}\right) /T_{c}$\ as functions of the lightest Higgs mass $m_{1}$ in
Fig. \ref{2}.

Comparing the number of points above and below the dash-dotted line in both
cases \textit{(a)} and \textit{(b)} in Fig. \ref{2}, we remark that the
first order phase transition is stronger than that of the Standard Model
with both criteria. However according to the large number of points below
the dash-dotted in \textit{(a)}, there are a lot of points which satisfy (%
\ref{con2}) but they do not really give a strong first order phase
transition according to (\ref{con1}).

When comparing the points in Fig. \ref{2}-a with the curve which represents
the Standard Model case, we remark that the addition of a singlet increases,
in general, the quantity $E_{Sp}(T_{c})/T_{c}$\ which is relevant to the
phase transition strength; that there are even a large number of points
above the line $E_{Sp}(T_{c})/T_{c}=45$.

The passage from the criterion (\ref{con1}), which is model-independent, to (%
\ref{con}), was based on two assumptions \cite{phicond}:

\textit{(I)} The sphaleron energy $E_{Sp}(T)$ scales like the vev $\upsilon
(T)$.\footnote{%
As mentioned above, this was verified for the Standard Model \cite{SMcheck};
and the Minimal Supersymmetric Standard Model \cite{MSSMcheck}.}

\textit{(II)} The sphaleron energy at $T=0$, is taken to be 1.87 in units of
$4\pi \upsilon /g$.

If the assumption \textit{(I)} is satisfied in our model 'SM+S', i.e. $%
E_{Sp}(T)\varpropto \Omega (T)$; and $E_{Sp}\left( 0\right) \simeq
1.87\times 4\pi \Omega \left( 0\right) /g$, then (\ref{con2}) is the
condition of a strong first order phase transition, but this not the case as
mentioned above.

In general, the value of the sphaleron energy at zero temperature is
significantly different from 1.87 in units of ($4\pi \Omega \left( 0\right)
/g$), thus if the assumption \textit{(I)} is fulfilled, then the criterion (%
\ref{con2}) is still viable but should be relaxed as $\Omega
_{c}/T_{c}\gtrsim 1+\delta $, where $\delta $\ describes the deviation from
the assumption \textit{(II)}.

\begin{figure}[t]
\begin{center}
\includegraphics[width=9cm,height=7cm]{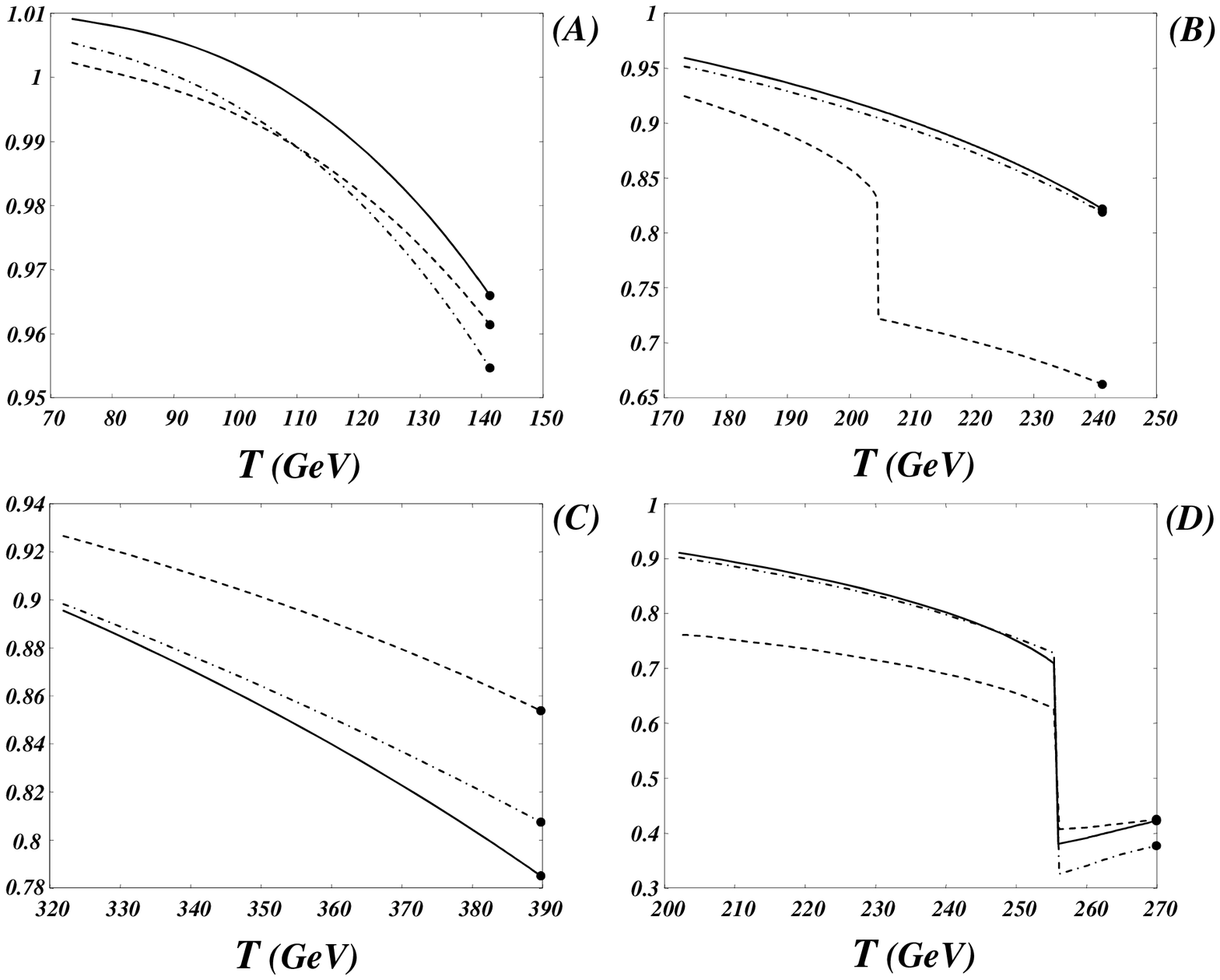}
\end{center}
\caption{\textit{The solid line denotes the ratio }$\protect\upsilon \left(
T\right) /\protect\upsilon $\textit{, the dashed one denotes }$\Omega \left(
T\right) /\Omega $\textit{; and the dot-dashed one denotes }$%
E_{Sp}(T)/E_{Sp}(0)$\textit{. All the plots end at the critical temperature.}
}
\label{3}
\end{figure}

In order to probe the assumption \textit{(I)} for our case, i.e.%
\begin{equation}
E_{Sp}\left( T\right) \varpropto \Omega \left( T\right) ,  \label{scalw}
\end{equation}%
we take the sets (A), (B), (C) and (D) used in table \ref{tab1} in the
previous section, and plot the ratios $\upsilon \left( T\right) /\upsilon
\left( 0\right) $, $\Omega \left( T\right) /\Omega \left( 0\right) $ and $%
E_{Sp}\left( T\right) /E_{Sp}\left( 0\right) $;\ as functions of
temperature, which lies between the critical temperature and another value.
The results are shown in Fig. \ref{3}.

Let us here comment on Fig. \ref{3}. For the case of (A), the ratio $E_{Sp}$(%
$T$)/$E_{Sp}$($0$) is close to both $\upsilon $($T$)/$\upsilon $($0$) and $%
\Omega $($T$)/$\Omega $($0$), which is almost 1 at $T_{c}$. For the case of
(B), the ratio $E_{Sp}$($T$)/$E_{Sp}$($0$) is very close to $\upsilon $($T$)/%
$\upsilon $($0$); however it is different a bit from 1 at $T_{c}$. At the
temperature $T\simeq 204.5$ $GeV$, there exists a secondary first order
phase transition, it happens on the axis $h=0$, where the false vacuum ($%
0,x_{0}$) is changed suddenly. In the case (C), the ratio $E_{Sp}$($T$)/$%
E_{Sp}$($0$) is closer to $\upsilon $($T$)/$\upsilon $($0$) than to $\Omega $%
($T$)/$\Omega $($0$); it is also different from 1 at $T_{c}$. In the last
case (D), there is also a secondary first order phase transition around $%
T\simeq 256$ $GeV$; where the true vacuum ($\upsilon ,x$) changes
discontinuously. We cannot call this an EWPT because the scalar $h$ has
already developed its vev. The ratio $E_{Sp}$($T$)/$E_{Sp}$($0$) is still
scaling like $\upsilon $($T$)/$\upsilon $($0$), but significantly different
from 1 at $T_{c}$.

It is clear that $E_{Sp}$($T$)\ does not scale like $\Omega $($T$), but
roughly speaking it scales like $\upsilon $($T$); with a little deviation in
some cases.

We claimed previously that the contribution of the singlet $S$ to the
sphaleron energy is small, and therefore this may be the reason why $%
E_{Sp}(T)$\ does not behave like $\Omega (T)$; and also does not behave
exactly like $\upsilon $($T$). In order to estimate the effect of the
singlet field $S$ on the sphaleron energy (\ref{enfuncs}), we compute the
sphaleron energy (\ref{enfuncs}) with replacing the singlet $S$ by its vev $%
x $, which we denote $\mathcal{E}_{Sp}(T)$. Then the fifth term in (\ref%
{enfuncs}) disappears and the third equation in (\ref{eqsms} and \ref{DifEq}%
) disappears also; the problem is reduced to a $SU(2)_{L}$ Higgs-gauge
system \cite{sphaleron}; but with a modified potential $\mathcal{V}_{eff}(h)$%
=$V_{eff}(h,x,T)$. We find that the singlet $S$ gives a negative
contribution to the sphaleron energy which is larger at higher temperatures.
But the contribution size is, generally, negligible. The maximum
contribution in the case (A) is less or equals 2.1 \%, less than 1.1 \% for
in case (B), and almost zero in the case (C): it is less than 0.08 \%; and
this is expected because the function R in Fig. \ref{1}-C is very close to
1. In the case (D), there are two different phases: at the first one before
the secondary phase transition i.e, $T_{c}>T>256$ GeV; the singlet
contribution is significant (between 8$\sim $9 \%), this may be due to the
smallness of the Higgs doublet vev at this range. While in the second phase $%
T<256$ GeV; the singlet contribution, as in the other cases, is less than 2
\%. Then in the absence of secondary first order phase transitions, we can
neglect the singlet contribution, but in its presence the singlet
contribution can be sizeable but not as large as that of the Higgs doublet
or gauge fields.

To justify this picture, we take again 3000 random sets of parameters and
plot $E_{Sp}(T_{c})/T_{c}$\ as a function of $\Omega _{c}/T_{c}$ in Fig. \ref%
{4}.

\begin{figure}[h]
\begin{center}
\includegraphics[width=7.5cm,height=5.5cm]{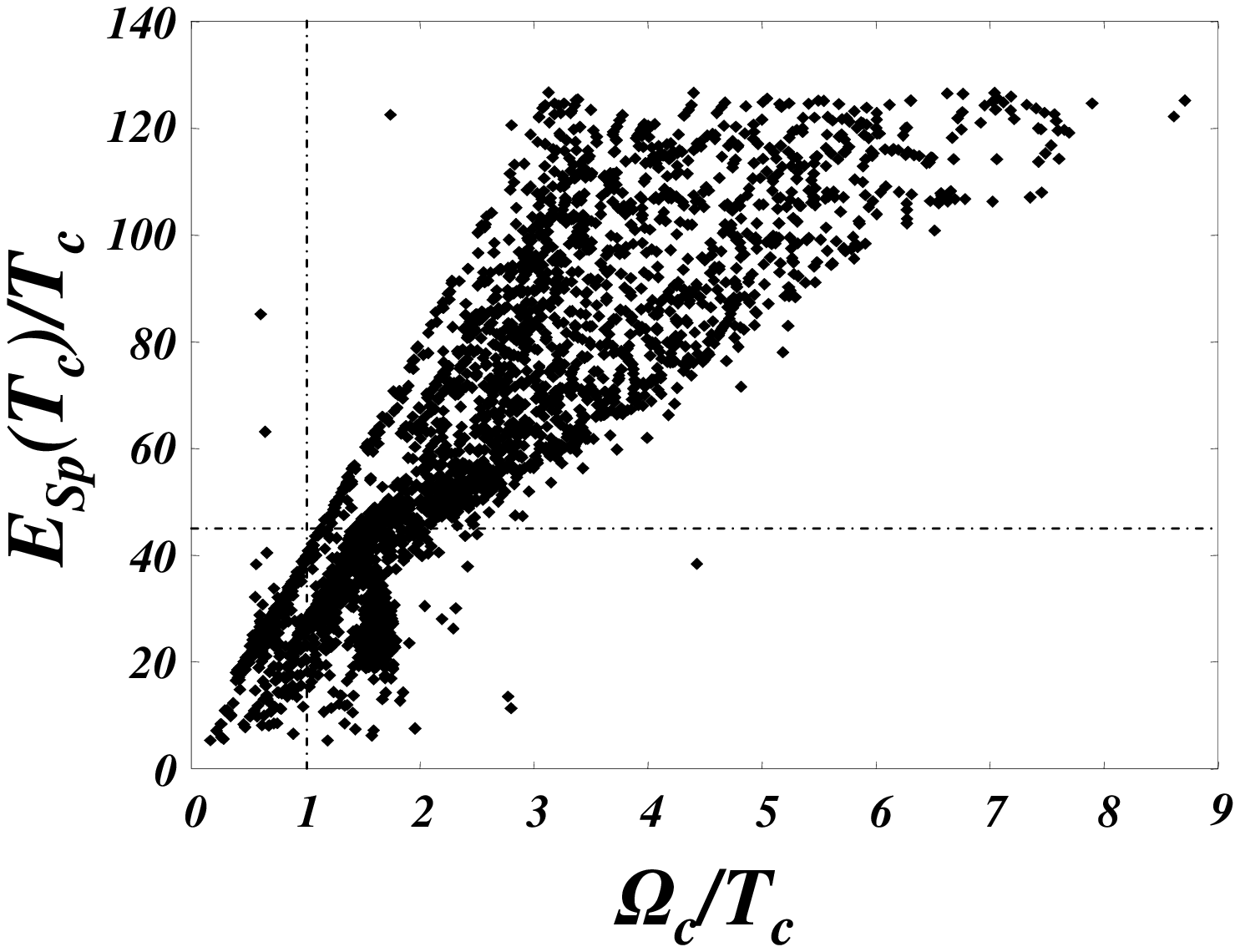}
\end{center}
\caption{$E_{Sp}(T_{c})/T_{c}$\textit{\textit{\ vs }}$\Omega _{c}/T_{c}$%
\textit{\textit{\ for 3000 randomly chosen sets of parameters.}}}
\label{4}
\end{figure}

Since there exist too many points in the region ($E_{Sp}(T_{c})/T_{c}\leq 45$
$\cap $ $\Omega (T_{c})/T_{c}\geq 1$), the criterion (\ref{con2}) is not the
definition of a strong first order EWPT. However it is satisfied for all
points that give really a strong first order EWPT except for 10 points due
to the existence of secondary first order phase transitions. Then we are now
sure that (\ref{con2})\ does not describe a strong first order EWPT.

In the sphaleron transitions, the singlet $S$ has no relation to lepton or
baryon number breaking phenomena. It does not couple to fermions or gauge
bosons; it is just a compensating field in the field equations; (\ref{eqsms}%
) and (\ref{DifEq}); and its effect on the sphaleron transition is
negligible as shown above. Then we claim that only the Higgs doublet vev is
relevant for the phase transition strength.

\begin{figure}[h]
\begin{center}
\includegraphics[width=7.5cm,height=5.5cm]{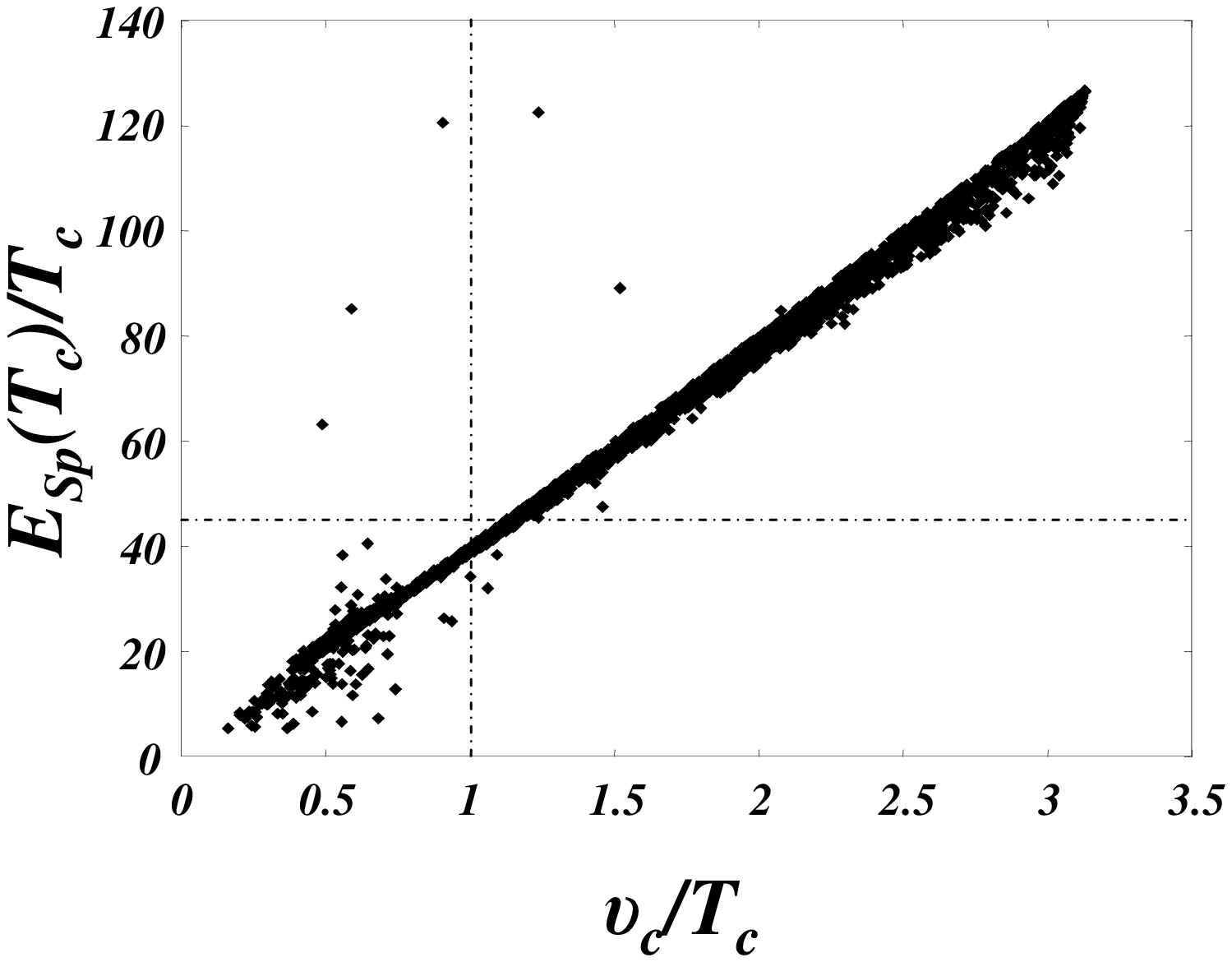}
\end{center}
\caption{$E_{Sp}(T_{c})/T_{c}$\textit{\textit{\ vs }}$\protect\upsilon %
_{c}/T_{c}$\textit{\textit{\ for 3000 randomly chosen sets of parameters.}}}
\label{5}
\end{figure}

We take 3000 random sets of parameters used previously, and plot $%
E_{Sp}(T_{c})/T_{c}$\ as a function of $\upsilon _{c}/T_{c}$ in Fig. \ref{5}.

It is clear that $E_{Sp}(T_{c})/T_{c}$ scales almost exactly\footnote{%
Except some points due to the existence of secondary first order phase
transitions; or due to the significant singlet contribution to the sphaleron
energy; especially for smaller Higgs vev values.} like $\upsilon _{c}/T_{c}$
except for some points, and (\ref{con}) can describe the strong first order
EWPT criterion for most of the points. Then when studying the EWPT in models
with a gauge singlet, one should treat the problem as the SM case (in case
of one doublet) with replacing the singlet by its vev; and look for the
Higgs vev in the path $\left. \partial V_{eff}(h,S)/\partial S\right\vert
_{S=x}=0$; whether it is larger than the critical temperature i.e, $\upsilon
_{c}/T_{c}\geq 1$?

With this modified potential $\mathcal{V}_{eff}(h)=\left.
V_{eff}(h,S)\right\vert _{S=x}$; the EWPT can be obtained easily as done by
the authors in \cite{SmS2}.

\section{Conclusion}

In this paper, the electroweak phase transition for the Standard Model with
a singlet is studied using the known criteria in the literature in addition
to the model-independent criterion found in \cite{cond}. The authors \cite%
{SmS,SmS0} found that the EWPT gets stronger even for Higgs masses above the
bound (\ref{mhbound}). They\ modified the simple criterion (\ref{con}) into (%
\ref{con2}), where they replaced the Higgs vev by the distance between the
two degenerate minima in the $h-S$ plan.

In our work, we checked whether this criterion is viable for this kind of
models or not. We took the Standard Model with a real singlet, then we
studied the EWPT using the sphaleron configuration at the critical
temperature, then we checked whether all the steps of the passage from the
model independent criterion (\ref{con1}) to (\ref{con}) in the Standard
Model case, are respected for our model (i.e. the passage from (\ref{con1})
to (\ref{con2})) or not?

We found that the EWPT gets stronger even for Higgs masses larger than 100
GeV; and this model does not suffer from the severe Higgs mass bound (\ref%
{mhbound}). However, we remarked that a sizeable number of the parameters
sets satisfy the modified criterion but do not really give a strong first
order EWPT, this allowed us to conclude that the 'modified criterion' is not
the criterion that describes a strong first order EWPT.

In order to understand why this modified criterion is not viable in this
case, we returned back to the SM to see how the passage from the
model-independent criterion to the simpler one proceeds? We found that the
two assumptions needed for the passage to the simpler criterion are not
fulfilled, in general, in our model 'SM+S':

$\bullet $ The sphaleron energy at zero temperature is different from the
value$\ 1.87$ in units of ($4\pi \Omega /g$).

$\bullet $ The sphaleron energy at finite temperature does not scale like $%
\Omega \left( T\right) $.

We guess that the reason for this is that the singlet does not couple to the
gauge field, then the missing of some contributions to the sphaleron energy
like $R^{2}\left( 1-f\right) ^{2}$ in (\ref{enfuncs}), can spoil the scaling
law, $E_{Sp}\left( T\right) \varpropto \Omega \left( T\right) $. This can be
inspired if we compare this situation with the case of the MSSM, where this
scaling law does work; and the general form of the sphaleron energy is
invariant under $h_{1}\leftrightarrow h_{2}$. The fact that the singlet does
couple only to the Higgs doublet leads to a singlet profile in the sphaleron
configuration in a Neumann type at $r=0$, which makes the singlet
contribution too small. Another important remark is that the possibility of
secondary first order phase transitions can, sometimes, spoil this scaling
law.

As a conclusion, we can say that the condition $\Omega _{c}/T_{c}\geq 1$\ is
not valid as a strongly first order phase transition criterion. But the
usual condition $\upsilon _{c}/T_{c}\geq 1$, is still the viable\ one, which
can describe the strong first order phase transition for the majority of the
physically allowed parameters as stated in Fig. \ref{5}. Moreover, this can
be satisfied even for Higgs masses in excess of 100 GeV unlike in the
Standard Model.

Then in such a model where the singlets couple only to the Higgs doublets,
it is convenient to study the EWPT within an effective model that contains
only doublets, where the singlets are replaced by their vev's. We expect
similar conclusion for models like the Next-to-Minimal Supersymmetric
Standard Model (NMSSM), where in this model the singlet couples only to the
two Higgs doublets; and its profile is a Neumann type in the sphaleron
configuration. Then the criterion for a strong first order EWPT is $\left\{
\upsilon _{1}^{2}+\upsilon _{2}^{2}\right\} ^{\frac{1}{2}}/T\geq 1$ at the
critical temperature, instead of $\left\{ \upsilon _{1}^{2}+\upsilon
_{2}^{2}+(x-x_{0})^{2}\right\} ^{\frac{1}{2}}/T\geq 1$.

\textbf{Acknowledgments}\newline

I want to thank Mikko Laine for useful remarks and corrections in the
manuscript. The author is supported by DFG.

\appendix

\section{The Boundary conditions}

To find the boundary conditions of (\ref{DifEq}), one should take into
account that the energy functional (\ref{enfuncs}) should be finite. It is
clear that in order for the contributions of the second and fourth terms in (%
\ref{enfuncs}) to be finite, $f$ must go to unity at the limit $\zeta
\rightarrow \infty $. According to the sphaleron definition, scalars go to
their vacuum at the infinity, i.e. $L,$ $R\rightarrow 1$ when $\zeta
\rightarrow \infty $, which makes the last term contribution to (\ref%
{enfuncs}) finite. Thus one can write all the functions as $1-c_{i}\exp
\left\{ -d_{i}\zeta \right\} $, and find the values of $c_{i}$ and $d_{i}$\
by inserting this behavior into the differential equations (\ref{DifEq}).

In the limit $\zeta \sim 0$, let us assume that the functions $f$, $L$ and $%
R $ have the profiles%
\begin{eqnarray}
f(\zeta ) &\sim &\zeta ^{n_{f}}  \notag \\
L(\zeta ) &\sim &c_{1}+\zeta ^{n_{L}}  \notag \\
R(\zeta ) &\sim &c_{2}+\zeta ^{n_{R}},  \label{profiles}
\end{eqnarray}%
where $n_{f}$, $n_{L}$ and $n_{R}$ are some positive constants. In this
limit, (\ref{DifEq}) can be approximated as%
\begin{eqnarray}
\frac{\partial ^{2}}{\partial \zeta ^{2}}f &\simeq &\frac{2}{\zeta ^{2}}f-%
\tfrac{1}{4}\frac{\upsilon ^{2}}{\Omega ^{2}}L^{2}  \notag \\
\frac{\partial ^{2}}{\partial \zeta ^{2}}L &\simeq &-\frac{2}{\zeta }\frac{%
\partial }{\partial \zeta }L+\frac{2}{\zeta ^{2}}L  \notag \\
\frac{\partial ^{2}}{\partial \zeta ^{2}}R &\simeq &-\frac{2}{\zeta }\frac{%
\partial }{\partial \zeta }R+\tfrac{1}{g^{2}x\Omega ^{2}}\left. \tfrac{%
\partial V_{eff}\left( h,S,T\right) }{\partial S}\right\vert _{\substack{ %
h=\upsilon L  \\ S=xR}}.  \label{eqappr}
\end{eqnarray}

From the second equation in (\ref{eqappr}), one can easily conclude that $%
L\sim \zeta $ or $\sim \zeta ^{-2}$, however the second choice makes the
energy functional integral (\ref{enfuncs}) divergent, thus $L\sim \zeta $ or
$\left\{ c_{1}=0,~n_{L}=1\right\} $. Using this result, one can conclude
from first equation in (\ref{eqappr}) that $f$ $\sim \zeta ^{2}$. However
the situation is different for the last equation (\ref{eqappr}), then one
can make%
\begin{eqnarray}
\tfrac{1}{g^{2}x\Omega ^{2}}\left. \tfrac{\partial V_{eff}\left(
h,S,T\right) }{\partial S}\right\vert _{\substack{ h=\upsilon L  \\ S=xR}}
&\sim &a\zeta ^{2}+\left\{ A+b\zeta ^{2}\right\} R\left( \zeta \right)
\notag \\
&&+BR^{2}\left( \zeta \right) +CR^{3}\left( \zeta \right) ,  \label{expvs}
\end{eqnarray}%
then inserting (\ref{profiles}) in (\ref{expvs}), one finds that the only
possibilities are $n_{R}=-1$ and $n_{R}=2$, where the first choice is
excluded in order that the energy functional integral (\ref{enfuncs}) to be
convergent, thus $R\sim a+b\zeta ^{2}$. Therefore at $\zeta =0$, $R$
satisfies the boundary condition of Neumann type, while $f$ and $L$ satisfy
those of Dirichlet type. The boundary conditions are summarized in (\ref{bc}%
).


\begin{thebibliography}{99}
\bibitem{BBN} B. Fields and S. Sarkar, astro-ph/0601514.

\bibitem{wmap} D. N. Spergel \textit{et al.}, astro-ph/0603449.

\bibitem{rev} A. Riotto, hep-ph/9807454.

\bibitem{sak} A.D. Sakharov, JETP Lett. \textbf{5}, 24 (1967).

\bibitem{ewbar} V.A. Rubakov and M.E. Shaposhnikov, Usp. Fiz. Nauk \textbf{%
166}, 493 (1996) [ Phys. Usp. \textbf{39}, 461 (1996).

\bibitem{thooft} G. 't Hooft, Phys. Rev. Lett. \textbf{37}, 8 (1976); Phys.
Rev. D \textbf{14}, 3432 (1976); \textbf{18}, 2199(E) (1978).

\bibitem{sphaleron} R.F. Klinkhamer and N.S. Manton, Phys. Rev.  D \textbf{30%
}, 2212 (1984).

\bibitem{cond} A.I. Bochkarev, S.V. Kuzmin and M.E. Shaposhnikov, Phys. Rev.
D \textbf{43}, 369 (1991).

\bibitem{SMcheck} S. Braibant, Y. Brihaye and J. Kunz, Int. J. Mod. Phys. A%
\textbf{8}, 5563 (1993).

\bibitem{MSSMcheck} J.M. Moreno, D.H. Oaknin, M. Quiros, Nucl. Phys. B%
\textbf{483}, 267 (1997).

\bibitem{phicond} M.E. Shaposhnikov, Nucl. Phys. B\textbf{287}, 757 (1987); B%
\textbf{299}, 797 (1988).

\bibitem{mhbound} A.I. Bochkarev and M.E. Shaposhnikov, Mod. Phys. Lett A%
\textbf{2}, 417 (1987).

\bibitem{Heb} Z. Fodor and A. Hebecker, Nucl. Phys. B\textbf{432}, 127
(1994).

\bibitem{LEP} Particle Data Group (W-M Yao et al.), J. Phys. G: Nucl. Part.
Phys. \textbf{33}, 1 (2006).

\bibitem{zvs} G.W. Anderson and L.J. Hall, Phys. Rev. D \textbf{45}, 2685
(1992); K.E.C. Benson, Phys. Rev. D \textbf{48}, 2456 (1993); J.R. Espinosa
and M. Quiros, Phys. Lett B\textbf{305}, 98 (1993).

\bibitem{SmS} J. Choi and R.R. Volkas, Phys. Lett. B\textbf{317}, 385
(1993); S.W. Ham, Y.S. Jeong and S.K. Oh, J. Phys. G\textbf{31}, 857 (2005).

\bibitem{SmS0} Y. Kondo, I. Umemura, K. Yamamoto, Phys. Lett. B\textbf{263},
93 (1991); N. Sei, I. Umemura and K. Yamamoto, Phys. Lett. B\textbf{299},
286 (1993).

\bibitem{fmixing} J. Kunz, B. Kleihaus and Y. Brihaye, Phys. Rev.  D \textbf{%
46}, 3587 (1992); B. Kleihaus, J. Kunz and Y. Brihaye, Phys. Lett. B\textbf{%
273}, 100 (1991).

\bibitem{thermal} L. Dolan and R. Jackiw, Phys. Rev.  D \textbf{9}, 3320
(1974).

\bibitem{ring} M.E. Carrington, Phys. Rev.  D 45, 2933 (1992).

\bibitem{spmod} T. Akiba, H. Kikuchi and T. Yanagida, Phys.~Rev.  D \textbf{%
38}, 1937 (1988); \textbf{40}, 588 (1989).

\bibitem{choi} J. Choi, Phys. Lett. B\textbf{345}, 253 (1995).

\bibitem{SpNMSSM} K. Funakubo, A. Kakuto, S. Tao and F. Toyoda, Prog. Theor.
Phys. \textbf{114}, 1069 (2005).

\bibitem{nr} W.H. Press, S.A. Teukolsky, W.T. Vetterling and B.P. Flannery,
\textit{Numerical Recipes in Fortran 77: The Art of Scientific Computing}
(Cambridge University Press, 1992).

\bibitem{SmS2} K. Funakubo, S. Tao and F. Toyoda, Prog. Theor. Phys. \textbf{%
114}, 369 (2005); M. Pietroni, Nucl. Phys. B\textbf{402}, 27 (1993); A.T.
Davies, C.D. Froggatt and R.G. Moorhouse, Phys. Lett. B\textbf{372}, 88
(1996); S.J. Huber and M.G. Schmidt, Nucl. Phys. B\textbf{606}, 183 (2001);
A. Menon, D.E. Morrissey and C.E.M. Wagner, Phys. Rev.  D \textbf{70},
035005 (2004).
\end{thebibliography}
\end{document}